\documentclass[letterpaper]{article} 
\usepackage[]{aaai2026}  
\usepackage{times}  
\usepackage{helvet}  
\usepackage{courier}  
\usepackage[hyphens]{url}  
\usepackage{graphicx} 
\usepackage{natbib}  
\usepackage{caption} 
\usepackage{algorithm}
\usepackage{algorithmic}
\usepackage{amsmath}
\usepackage{amsfonts}

\usepackage{multirow}
\usepackage[table,xcdraw]{xcolor}
\usepackage{booktabs}
\usepackage{newfloat}
\usepackage{listings}
\floatstyle{ruled}
\newfloat{listing}{tb}{lst}{}
\floatname{listing}{Listing}
\title{Perceive-Sample-Compress: Towards Real-Time 3D Gaussian Splatting}
\author {
    Zijian Wang \thanks{These authors contributed equally to this work.}  \textsuperscript{\rm 1},
    Beizhen Zhao \footnotemark[1]  \textsuperscript{\rm 1},
    Hao Wang  \thanks{Corresponding author.}  \textsuperscript{\rm 1}
}
\affiliations {
    \textsuperscript{\rm 1} ‌The Hong Kong University of Science and Technology (Guangzhou) \\
    zwang886@connect.hkust-gz.edu.cn, bzhao610@connect.hkust-gz.edu.cn, haowang@hkust-gz.edu.cn
}

\usepackage{bibentry}

\begin{document}

\maketitle

\begin{abstract}
Recent advances in 3D Gaussian Splatting (3DGS) have demonstrated remarkable capabilities in real-time and photorealistic novel view synthesis. 
However, traditional 3DGS representations often struggle with large-scale scene management and efficient storage, particularly when dealing with complex environments or limited computational resources. 
To address these limitations, we introduce a novel perceive-sample-compress framework for 3D Gaussian Splatting. 
Specifically, we propose a scene perception compensation algorithm that intelligently refines Gaussian parameters at each level.
This algorithm intelligently prioritizes visual importance for higher fidelity rendering in critical areas, while optimizing resource usage and improving overall visible quality.
Furthermore, we propose a pyramid sampling representation to manage Gaussian primitives across hierarchical levels. 
Finally, to facilitate efficient storage of proposed hierarchical pyramid representations, we develop a Generalized Gaussian Mixed model compression algorithm to achieve significant compression ratios without sacrificing visual fidelity.
The extensive experiments demonstrate that our method significantly improves memory efficiency and high visual quality while maintaining real-time rendering speed.

\end{abstract}

\section{Introduction}

3D Gaussian Splatting (3DGS)~\cite{kerbl20233d} has revolutionized the field of novel view synthesis (NVS)~\cite{avidan1997novel, watson2022novel, choi2019extreme, riegler2020free, dalal2024gaussian}, offering a compelling solution to traditional volumetric rendering and mesh-based methods~\cite{yariv2023bakedsdf, lombardi2019neural, penner2017soft, wizadwongsa2021nex, zhou2018stereo}. 
By representing scenes as a set of anisotropic Gaussians, 3DGS achieves remarkable photorealism and real-time rendering speeds, quickly becoming a cornerstone technology for 3D reconstruction. 

Despite its impressive capabilities, a key challenge lies in managing the computational complexity and storage requirements when scenes involve millions of Gaussian points. 
The direct application of 3DGS to large-scale or memory-constrained scenarios lacks efficient management representation~\cite{wang2024pygs, liu2024citygaussian, chen2024gigags}, resulting in inconsistencies and degradation when rendering scenes at different scales. 
When scenes involve millions of Gaussian primitives, the dense collection of Gaussians can lead to excessive memory consumption and computational demands, hindering its scalability.

Existing methods struggle to effectively address the challenges of storage consumption and hierarchical scalability simultaneously. 
Some attempts to address the storage consumption of Gaussian Splatting, such as Scaffold-GS~\cite{lu2024scaffold}, HAC++~\cite{chen2025hac++} and Context-GS~\cite{wang2024contextgs}, lack explicit Level-of-Detail (LOD)~\cite{david2003level} mechanisms. 
This absence results in inefficient storage and rendering, as the entire scene representation must be processed regardless of the viewer’s proximity or the required level of detail. 
Other approaches, like Octree-GS~\cite{ren2024octree}, while introducing hierarchical structures, can be overly dependent on camera distance to optimize the whole structure and may lead to suboptimal performance in scenes with complex geometric distributions, as they might over-prune crucial details and fail to capture fine-grained information.

To overcome these limitations, we introduce a novel perceive-sample-compress framework for 3D Gaussian Splatting. 
We propose a scene perception compensation algorithm that refines the parameters at each level. 
Unlike existing methods that rely solely on camera distance, our algorithm dynamically prioritizes optimization efforts based on camera coverage area and depth of field. 
This algorithm intelligently prioritizes visual importance for higher fidelity rendering in critical areas, while optimizing resource usage and improving overall visible quality.

Then, based on the levels of perception areas, we propose a pyramid sampling representation inspired by the Laplacian pyramid and voxelization.
This approach allows us to group and represent Gaussians at multiple scales for each level of the pyramid. 
By voxelizing the scene and applying pyramid to aggregate Gaussian properties within these voxels, we establish a hierarchical multi-scale scene structure. 
This pyramid representation significantly enhances memory efficiency and enables adaptive rendering strategies.

Finally, we propose a generalized Gaussian mixed model compression algorithm based on the hierarchical pyramid structure. 
By analyzing the distribution of Gaussian attributes across hierarchical levels, we can effectively encode redundant information and reduce the model size, making our framework more practical for storage and transmission.

In summary, our work makes three key contributions:

\begin{itemize}
    \item We introduce a scene perception compensation algorithm that intelligently refines Gaussian parameters at each level based on camera coverage area and depth of field, leading to improved visual quality.
    \item We propose a pyramid sampling function for 3D Gaussian Splatting by combining Laplacian pyramid with voxelization, enabling efficient hierarchical representation and adaptive processing of large-scale scenes.
    \item We develop a compression algorithm based on generalized Gaussian distribution for efficient storage of our Gaussian representations, significantly reducing memory consumption and facilitating broader applicability.
\end{itemize}

\section{Related Work}

\subsection{Level-of-Detail 3DGS}
Level-of-Detail (LOD)~\cite{david2003level} is a classic and critical technique in computer graphics that effectively manages the rendering budget of 3D scenes to improve rendering efficiency. 
Several works have explored LOD mechanisms in Gaussian-based scene representations~\cite{cui2024letsgo, seo2024flod, milef2025learning, ren2024octree, wang2024pygs, liu2024citygaussian, kerbl2024hierarchical}. 
For example, LetsGo~\cite{cui2024letsgo} employs a multi-resolution Gaussian model and jointly optimizes multiple levels. 
However, the rendering quality is highly dependent on multi-resolution point cloud inputs, which significantly increases training overhead.
CityGaussian~\cite{liu2024citygaussian} improves efficiency of large-scale scenes by selecting and blending LOD levels based on distance intervals, but requires manual setting of distance thresholds, which results in a lack of robustness. 
While Octree-GS~\cite{ren2024octree} significantly improves rendering speed through its use of an explicit octree structure and a cumulative LOD strategy, its level selection strategy relies excessively on camera distance, making it prone to losing critical visual details due to improper level switching in complex and heavily occluded scenes.

\begin{figure}[ht]
\centering
\includegraphics[width=0.4\textwidth]{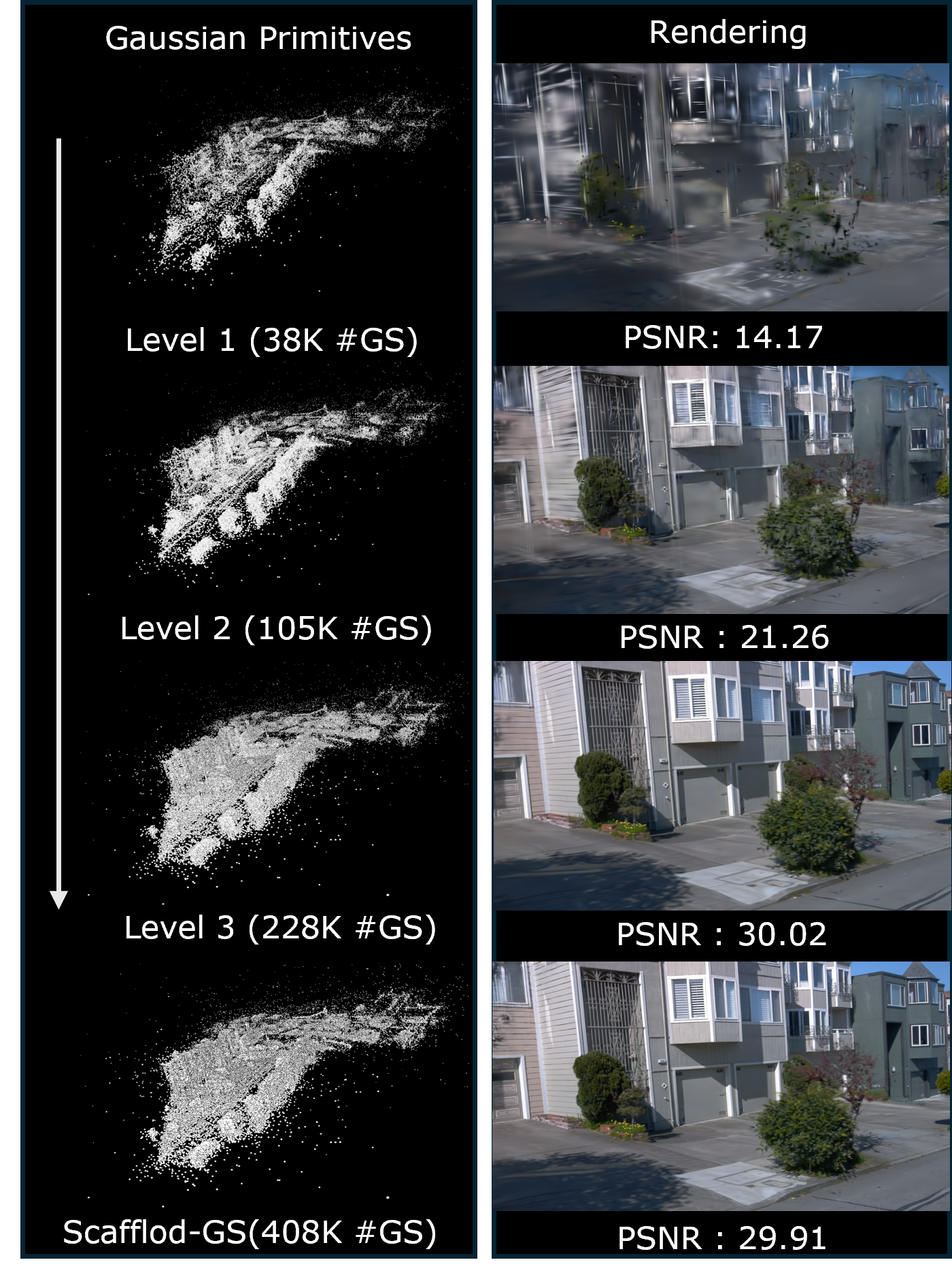}
\caption{\textbf{Visualization of pyramid level} Compared with Scaffold-GS, our method can achieve higher PSNR scores with fewer Gaussian primitives.}
\label{pyramid}
\end{figure}

\subsection{Compression of 3DGS}
Although 3DGS~\cite{kerbl20233d} demonstrates outstanding performance in terms of rendering speed and image fidelity, the large number of Gaussians and their associated parameters result in significant storage overhead. 
To address this issue, researchers have proposed various compression strategies.~\cite{lee2024compact, niedermayr2024compressed, girish2024eagles, fan2024lightgaussian, navaneet2023compact3d, papantonakis2024reducing, ali2024trimming, wang2024contextgs, liu2024compgs, chen2024hac, chen2025hac++} 
For example, the work in~\cite{niedermayr2024compressed, navaneet2023compact3d} introduces codebooks to cluster Gaussian parameters, while the studies in~\cite{papantonakis2024reducing, ali2024trimming} systematically explore pruning-based optimization methods.
Building on Scaffold-GS~\cite{lu2024scaffold}, Context-GS~\cite{wang2024contextgs} and CompGS~\cite{liu2024compgs} adopt context-aware approaches that explicitly incorporate hierarchical relationships among anchors and Gaussians to enhance representational capacity. 
HAC++~\cite{chen2025hac++} further introduces hashed features as priors for entropy coding and demonstrates its effectiveness in improving compression efficiency. 
However, these methods heavily rely on the anchor selection mechanism proposed in Scaffold-GS, and fail to fully model the distributional differences among various Gaussian parameters during encoding stage.

\begin{figure*}[t]
\centering
\includegraphics[width=0.95\textwidth]{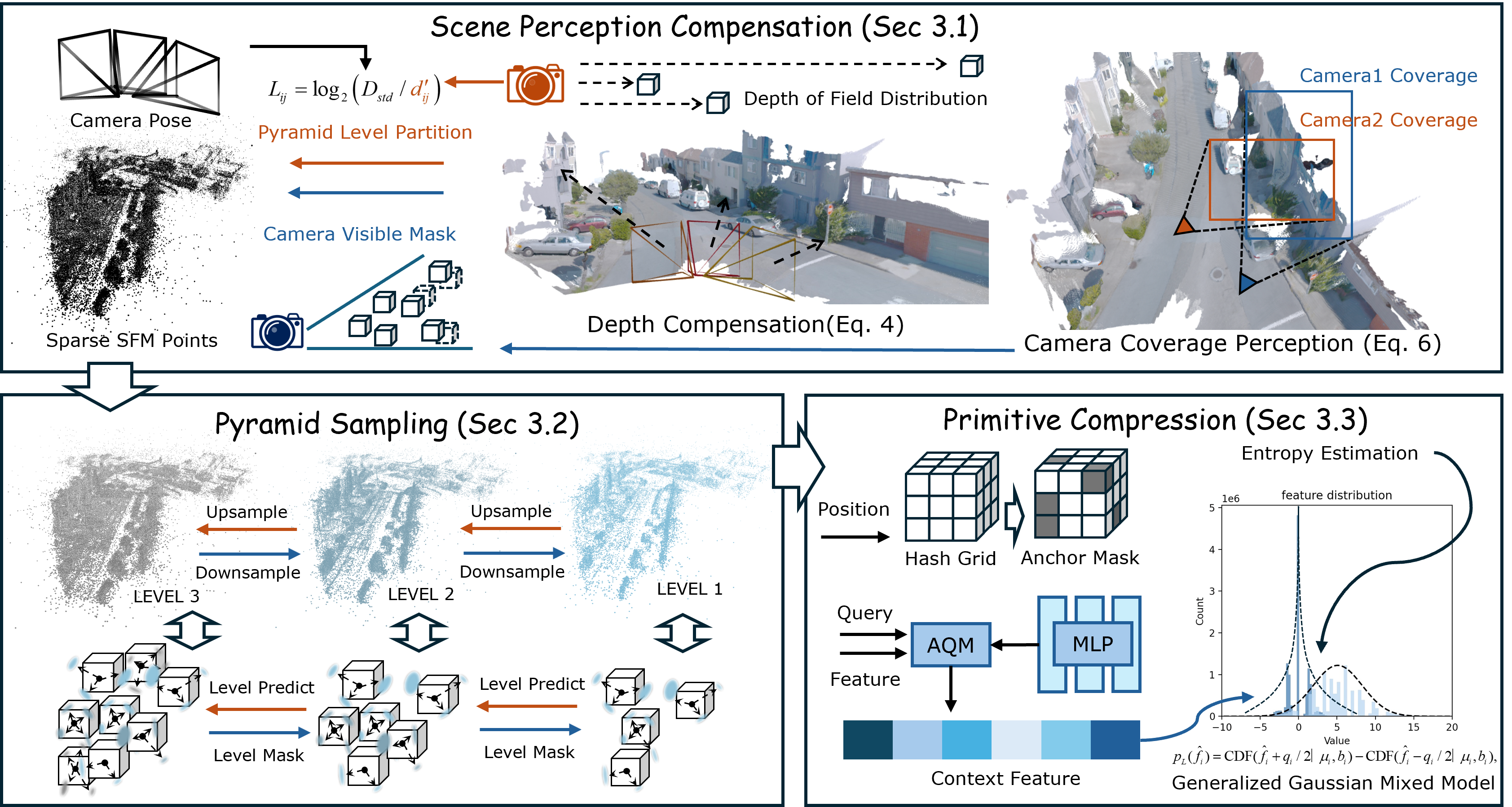}
\caption{\textbf{Framework of Pyramid-GS.} We begin by sampling the input sparse point cloud. 
We propose a scene perception compensation algorithm, which take the camera coverage and depth of field accounts for pyramid level partition.
Then we design a hierarchical pyramid sampling representation to set up pyramid structure.
Finally, we utilize compression algorithm for efficient storage while maintaining real-time rendering speed and photorealistic visual effects.
An adaptive quantization module (AQM) is used to generate context feature, while the generalized gaussian mixed model is used for entropy estimation.}
\label{framework}
\end{figure*}

\section{Methodology}

Our proposed approach aims to address the limitations of traditional 3DGS by introducing a robust pyramid structure framework. 
This framework is built upon three interconnected components: 
a scene perception compensation algorithm that leverages camera coverage and depth of field for improved rendering quality,
a hierarchical pyramid sampling representation with voxelization for efficient scene management,  
and a compression scheme for effective model storage. 
These components interact seamlessly, with the pyramid sampling establishing the foundation for multi-scale detail, the camera-driven compensation refining pyramid level partition accuracy, and the compression scheme ensuring scalability for large scenes.
The overall goal is to create a scalable, efficient, and high-fidelity 3D representation. 
This section details the technical implementation of these components.
The overall pipeline is shown in Fig.~\ref{framework}.

\subsection{Scene Perception Compensation}
\label{subsec:compensation_algorithm}

To refine the pyramid level partition and ensure perceptual quality, we introduce a compensation algorithm that considers camera coverage and depth of field (DOF) effects. 
Our algorithm aims to dynamically adjust the perceived importance of anchors based on their visibility and how they are affected by the camera's optical properties. 
We model this by computing a visibility score for each anchor, which incorporating both camera coverage and depth perception.

We approximate the DOF effect by analyzing the distribution of points along the principal viewing direction from the camera. 
For a set of anchors $\{A_k\}$ with position $\mathbf{p}_k \in \mathbb{R}^3$, we can estimate the variance of their projected distances from the camera. 
Let $\mathbf{d}_k = \mathbf{p}_k - \mathbf{c}_j$ be the vector from the camera to the $k$-th anchor. 
We identify the main axis of dispersion $\mathbf{m}_j$ for these points. 
The projected depth for $A_k$ is $z_k = |\mathbf{d}_k \cdot \mathbf{m}_j|$. We then compute the standard deviation of these projected depths, where $\sigma_{z,j} = \text{std}(z_k)$.

A large $\sigma_{z,j}$ indicates that many points are spread out in depth, suggesting a shallow DOF or a scene with significant depth variation. 
In such cases, we apply a depth compensation factor $f_{depth}$ to the distance used for pyramid level partition. 
This factor is designed such that a larger $\sigma_{z,j}$ leads to a smaller $f_{depth}$, effectively making objects appear "closer" in terms of pyramid priority, thus encouraging their representation at finer levels. 
The formulation is as follows:
\begin{equation}
f_{depth} = 1 + \alpha \cdot \max(0, \frac{\sigma_{z,j}}{\sigma_{z,thresh}} - 1),
\end{equation}
where $\alpha$ is a superparameter and $\sigma_{z,thresh}$ is a threshold for DOF significance. The adjusted distance $d'_{ij}$ for pyramid level partition for anchor $A_i$ from camera $c_j$ becomes $d'_{ij} = d_{ij} \cdot f_{depth}$, where $d_{ij} = \|\mathbf{p}_i - \mathbf{c}_j\|$.
The pyramid level $L_{ij}$ for $A_i$ from camera $c_j$ is then determined based on this adjusted distance, typically using a logarithmic scale:
\begin{equation}
L_{ij} = \log_{2} \left( \frac{D_{std}}{d'_{ij}} \right),
\end{equation}
where $D_{std}$ is a standard reference distance. This value is then mapped to an integer pyramid level.

For each camera $c_j$ with center $\mathbf{c}_j$ and intrinsic parameters, we first determine the visible anchors. 
An anchor is considered visible to camera $c_j$ if its predicted pyramid level $L_{ij}$ is smaller or equal to the current level.
We then count the number of cameras $N_{vis, i}$ that render at least one anchor at current level. 
The normalized coverage score for $G_i$ is 
\begin{equation}
    C_i = \beta \frac{N_{vis, i}}{N}.
\end{equation}

Then we calculate the mean value of the counts of visible anchors for each camera as the visible threshold $\tau$. 
The final visible threshold is updated through the coverage score $C$.
We can utilize this threshold for camera visibility control through a mask for scene perception.
\begin{equation}
    \tau_{new} = (1 + C) \tau_{old}.
\end{equation}

This scheme ensures that points with larger depth variation, often corresponding to distant or complex regions, are allocated appropriate levels to maintain visual quality. 
The method enhances the initialization and refinement of pyramid levels, leading to more accurate scene representations.

\subsection{Pyramid Sampling Representation}
\label{sec:pyramid}

To enable efficient management of Gaussian primitives, we introduce a hierarchical representation inspired by Laplacian pyramid and voxelization. 
Traditional methods often lack explicit level control, leading to uniform processing of all Gaussians, which is inefficient for large scenes. Our approach tackles this by organizing Gaussians into a multi-scale structure.
To effectively represent the scene at multiple levels of detail, we employ a Laplacian pyramid framework that decomposes the scene into several resolution layers. 
At each level, scene features are captured with varying degrees of detail, enabling scalable rendering and processing.

Given an input point cloud $\mathcal{P} = \{p_i \in \mathbb{R}^3\}_{i=1}^N$, we construct a Laplacian pyramid with $L$ levels through iterative voxelization.
First, we recursively group voxels into smaller blocks. 
At each level of the pyramid, Gaussians within a block are voxelized into representative anchors. 
This aggregation is performed using principles akin to the Laplacian pyramid construction, where each level captures the high-frequency details relative to the coarser level below:

\begin{equation}
\mathcal{V^{'}}_{l-1} = \mathcal{R}_{l-1} \cap \mathcal{P}
, \quad 
\mathcal{R}_{l-1} = \text{Downsample}(\mathcal{V^{'}}_{l}, \rho_{l-1}),
\end{equation}

where $\mathcal{V}_l$ denotes the voxelized points at level $l$ with resolution $\rho_l = 2^{-l}\rho_0$, and $\mathcal{R}_l$ stores residuals. 
Through the downsample process, we get a simple multi-scale structure representation.
Then we execute an upsample process to sample detail residuals of each level.

\begin{equation}
\mathcal{V}_{l} = \mathcal{R}_{l} \cup \mathcal{V^{'}}_l
, \quad 
\mathcal{R}_{l} = \text{Upsample}(\mathcal{V^{'}}_{l-1}, \rho_{l}).
\end{equation}

The core idea is to maintain a set of Gaussians that are adaptively represented across different spatial resolutions. 
The coarsest level represents the overall scene structure, while finer levels provide increasing detail. 
This allows us to selectively process parts of the scene at an appropriate resolution based on the viewing context. 
This voxel-based, pyramidical representation significantly reduces the number of active Gaussians and improves memory efficiency compared to a flat representation.
After voxelizing $\mathcal{P}$ into a grid $\mathcal{V}$, each voxel is parameterized by Gaussian functions predicted through a multi-layer perceptron (MLP):
\begin{equation}
\mathcal{G}_{l}(\mathbf{x}) = 
\left\{ 
F^{\sigma}(\mathbf{x}_{l}), 
F^{\mu}(\mathbf{x}_{l}), 
F^{\alpha}(\mathbf{x}_{l}), 
F^{c}(\mathbf{x}_{l})
\right\},
\end{equation}
where $\mathcal{G}_{l}$ denotes the Gaussian representation at level $l$. \( F^{\sigma} \), \( F^{\mu} \), \( F^{\alpha} \) and \( F^{c} \) represent the MLP network to generate variance $\sigma_{l}$, mean $\mu_{l}$, opacity ${\alpha}$ and color $c$ of the gaussians.

This multi-scale decomposition allows us to represent scene details hierarchically, such that coarse layers capture the global structure and finer layers preserve details—enabling efficient rendering and bias towards important scene features at different pyramid levels.

\subsection{Generalized Gaussian Mixed Compression}

To ensure the scalability of Pyramid-GS, we propose an efficient model compression strategy based on context-aware entropy coding, targeting anchor features across pyramid levels. Drawing inspiration from HAC++~\cite{chen2025hac++}, which employs hash-grid features to assist an Adaptive Quantization Module (AQM) for Gaussian parameter prediction, we introduce a key modification to the probability model to better capture the statistical characteristics of our hierarchical representation.

In image and video compression, the Generalized Gaussian Distribution (GGD) effectively models transform or prediction residuals. 
We adopt GGD to uniformly characterize diverse parameter distributions in our framework. Notably, high-frequency components in the pyramid exhibit sparsity, with distributions sharply peaked at zero and heavy tails, aligning with the Laplace distribution (GGD with $\beta=1$). 
In contrast, other parameters resemble Gaussian distributions (GGD with $\beta=2$), as illustrated in Fig.~\ref{framework}.
GGD offers a statistically sound foundation, enabling the application of tailored priors to different parameters within the framework. 
Its probability density function (PDF) is defined as:
\begin{equation}
\label{eq:ggd_pdf}
p(x | \mu, \alpha, \beta) = \frac{\beta}{2\alpha\Gamma(1/\beta)} \exp\left(-\left(\frac{|x - \mu|}{\alpha}\right)^\beta\right),
\end{equation}
where $\mu$ is the location parameter, $\alpha$ is the scale parameter, and $\beta$ is the shape parameter that controls the distribution's tail behavior.
Following HAC++~\cite{chen2025hac++}, we use a lightweight MLP, denoted as MLPc, to predict the probability distribution parameters for a given Gaussian attribute $\hat{f}^i$ from its spatial hash-grid feature $f^i_h$. 
Our MLPc estimates the location $\mu_i$ and scale $\alpha_i$ parameters for the distribution. The shape parameter $\beta_i$, however, is pre-set according to the type of parameter being compressed: for high-frequency residual features, we set $\beta_i=1$ (Laplace), while for other parameters, we set $\beta_i=2$ (Gaussian).

The probability of the quantized attribute $\hat{f}^i$ with a quantization step $q_i$ is then calculated by integrating the corresponding GGD's PDF over the quantization bin:
\begin{equation}
\label{eq:ggd_prob}
\begin{aligned}
p_{\text{GGD}}(\hat{f}^i) ={} & \ \text{CDF}(\hat{f}^i + q_i/2 \mid \mu_i, \alpha_i, \beta_i) \\
& - \text{CDF}(\hat{f}^i - q_i/2 \mid \mu_i, \alpha_i, \beta_i),
\end{aligned}
\end{equation}
where $\text{CDF}(x | \mu, \alpha, \beta)$ is the cumulative distribution function of the GGD, with $\mu_i, \alpha_i = \text{MLPc}(f^i_h)$ and $\beta_i$ being the pre-set value. This design allows the model to leverage the generality of the GGD framework to impose the most fitting inductive bias on data with different statistical characteristics, thereby maximizing compression efficiency.

This compression module is applied to the attributes of Gaussians at each level of our pyramid. The entire model is trained end-to-end by minimizing a rate-distortion loss, which balances rendering quality and model size:
\begin{equation}
\label{eq:loss_function}
\mathcal{L} = \mathcal{L}_{\text{render}} + \lambda \left( \frac{1}{N} \sum_{i=1}^{N} -\log_2 p_{\text{GGD}}(\hat{f}^i) + \mathcal{L}_{\text{hash}} \right).
\end{equation}
Here, $\mathcal{L}_{\text{render}}$ combines fidelity losses akin to Scaffold-GS, while the second term denotes the overall rate loss, comprising the average entropy of primitive attributes and the hash-grid entropy. With $N$ as the number of primitives and $\lambda$ controlling the quality–compression trade-off, this formulation encourages compact yet high-fidelity representations.

\section{Experiment}
In this section, we begin by outlining the datasets used for our evaluation and providing the specific implementation details of our experiments.
Then we present our main comparison results and the ablation study and related discussions.

\begin{table*}[hp]
  \centering
  \caption{Quantitative comparison on Waymo~\cite{sun2020scalability} and MatrixCity~\cite{li2023matrixcity} datasets. The color of each cell shows the \colorbox[HTML]{F09BA0}{best} and the \colorbox[HTML]{FCCF93}{second best}.}
  \label{tab:comparison_metrics_waymo_matrix}
  \resizebox{0.95\textwidth}{!}{%
    \begin{tabular}{l|ccccc|ccccc}
      \toprule
      \multirow{2}{*}{\raisebox{-0.5\height}{\shortstack{Dataset\\Method}}} & \multicolumn{5}{c|}{Waymo~\cite{sun2020scalability}} & \multicolumn{5}{c}{MatrixCity~\cite{li2023matrixcity}} \\
      \cmidrule(lr){2-6} \cmidrule(lr){7-11}
      & SSIM↑ & PSNR↑ & LPIPS↓ & \#GS(k)↓ & Mem(MB)↓ & SSIM↑ & PSNR↑ & LPIPS↓ & \#GS(k)↓ & Mem(MB)↓ \\
      \midrule
      3DGS~\cite{kerbl20233d}          & 0.840 & 27.19 & 0.313 & 1530 & 382.6 & 0.823 & 26.82 & 0.246 & 1432 & 3387.4 \\
      2DGS~\cite{huang20242d}          & 0.801 & 24.83 & 0.373 & 860  & 187.1 & 0.818 & 26.28 & 0.232 & 832  & 1659.1 \\
      Hierarchical-GS~\cite{kerbl2024hierarchical}   & 0.808 & 24.85 & 0.307 & 530 & 1035.7 & 0.823 & 27.69 & 0.276 & 271 & 1866.7 \\
      Scaffold-GS~\cite{lu2024scaffold} & 0.843 & 27.51 & \cellcolor[HTML]{FCCF93}0.310 & 175  & 104.2 & 0.868 & 29.00 & 0.210 & 357  & 371.2 \\
      Octree-GS~\cite{ren2024octree}  & 0.828 & 26.82 & 0.321 & 171  & 97.5  & 0.887 & \cellcolor[HTML]{FCCF93}29.83 & 0.192 & 360  & 380.3 \\
      ContextGS~\cite{wang2024contextgs} & \cellcolor[HTML]{FCCF93}0.847 & \cellcolor[HTML]{FCCF93}27.93 & \cellcolor[HTML]{FCCF93}0.310 & 133  & \cellcolor[HTML]{FCCF93}11.1  & 0.878 & 29.26 & 0.188 & \cellcolor[HTML]{FCCF93}221  & \cellcolor[HTML]{F09BA0}30.3 \\
      HAC++~\cite{chen2025hac++}       & 0.846 & 27.68 & 0.311 & \cellcolor[HTML]{FCCF93}124  & 15.7  & \cellcolor[HTML]{FCCF93}0.884 & 29.29 & \cellcolor[HTML]{FCCF93}0.175 & \cellcolor[HTML]{F09BA0}206  & \cellcolor[HTML]{FCCF93}39.5 \\
      \midrule
      Ours                            & \cellcolor[HTML]{F09BA0}0.851 & \cellcolor[HTML]{F09BA0}28.18 & \cellcolor[HTML]{F09BA0}0.300 & \cellcolor[HTML]{F09BA0}105  & \cellcolor[HTML]{F09BA0}9.5   & \cellcolor[HTML]{F09BA0}0.896      &\cellcolor[HTML]{F09BA0}30.08       & \cellcolor[HTML]{F09BA0}0.154      &406       & 44.0      \\
      \bottomrule
    \end{tabular}
  }
\label{e1}
\end{table*}

\subsection{Dataset}
To evaluate the performance and scalability of our model, we selected 17 scenes from four widely-used datasets.

\begin{table*}[hp]
  \centering
  \caption{Quantitative comparison on Mip-NeRF360~\cite{barron2022mip} and Tanks $\&$ Temples datasets~\cite{knapitsch2017tanks}. The color of each cell shows the \colorbox[HTML]{F09BA0}{best} and the \colorbox[HTML]{FCCF93}{second best}.}
  \label{tab:comparison_metrics_mip_tt}
  \resizebox{0.95\textwidth}{!}{%
    \begin{tabular}{l|ccccc|ccccc}
      \toprule
      \multirow{2}{*}{\raisebox{-0.5\height}{\shortstack{Dataset\\Method}}} & \multicolumn{5}{c|}{Mip-NeRF360~\cite{barron2022mip}} & \multicolumn{5}{c}{Tanks $\&$ Temples~\cite{knapitsch2017tanks}} \\
      \cmidrule(lr){2-6} \cmidrule(lr){7-11}
      & SSIM↑ & PSNR↑ & LPIPS↓ & \#GS(k)↓ & Mem(MB)↓ & SSIM↑ & PSNR↑ & LPIPS↓ & \#GS(k)↓ & Mem(MB)↓ \\
      \midrule
      3DGS~\cite{kerbl20233d}        & \cellcolor[HTML]{FCCF93}0.870 & 28.69 & \cellcolor[HTML]{F09BA0}0.182 & 962 & 754.7 & 0.844 & 23.69 & 0.178 & 765 & 430.1 \\
      2DGS~\cite{huang20242d}        & 0.863 & 28.53 & 0.201 & \cellcolor[HTML]{FCCF93}399 & 390.4 & 0.830 & 23.25 & 0.212 & \cellcolor[HTML]{FCCF93}352 & 204.4 \\
      Scaffold-GS~\cite{lu2024scaffold} & \cellcolor[HTML]{FCCF93}0.870 & \cellcolor[HTML]{FCCF93}29.35 & 0.188 & 658 & 182.6 & 0.853 & 23.96 & 0.177 & 626 & 167.5 \\
      Octree-GS~\cite{ren2024octree}   & 0.867 & 29.11 & 0.188 & 695 & 140.4 & \cellcolor[HTML]{F09BA0}0.866 & \cellcolor[HTML]{F09BA0}24.68 & \cellcolor[HTML]{F09BA0}0.153 & 443 & 88.5 \\
      ContextGS~\cite{wang2024contextgs}   & 0.868 & 29.30 & 0.194 & 685 & 19.9  & \cellcolor[HTML]{FCCF93}0.855 & 24.29 & 0.176 & 469 & 11.8 \\
      HAC++~\cite{chen2025hac++}       & 0.865 & 29.26 & 0.207 & 680 & \cellcolor[HTML]{FCCF93}16.0  & 0.854 & \cellcolor[HTML]{FCCF93}24.32 & 0.178 & 481 & \cellcolor[HTML]{F09BA0}8.6 \\
      \midrule
      Ours        & \cellcolor[HTML]{F09BA0}0.876 & \cellcolor[HTML]{F09BA0}29.39 & \cellcolor[HTML]{FCCF93}0.187 & \cellcolor[HTML]{F09BA0}377 & \cellcolor[HTML]{F09BA0}15.8  & 0.847      & \cellcolor[HTML]{FCCF93}24.32     &  \cellcolor[HTML]{FCCF93}0.174     &\cellcolor[HTML]{F09BA0}317       & \cellcolor[HTML]{FCCF93}11.7       \\
      \bottomrule
    \end{tabular}
  }
  \label{e2}
\end{table*}

\paragraph{Small-Scale Datasets.}
To evaluate rendering fidelity on complex, object-centric scenes, we use two standard datasets.
Mip-NeRF 360~\cite{barron2022mip} contains seven challenging scenes featuring unbounded 360° camera trajectories, with approximately 200 images per scene. 
Tanks $\&$ Temples~\cite{knapitsch2017tanks} is used for assessing the reconstruction of fine-grained geometric details, each scene containing over 200 images.

\paragraph{Large-Scale Datasets.}
To demonstrate the scalability in large-scale environments, we employ two distinct datasets.
Waymo Open Dataset~\cite{sun2020scalability} provides real-world autonomous driving scenarios. 
We utilize six sequences, each with around 600 frames, to test our model's ability to handle extensive environments and long camera paths.
Matrix City~\cite{li2023matrixcity} is a massive, city-scale dataset that tests the limits of scalability. 
Following Octree-GS~\cite{ren2024octree}, we test on a blocksmall partition which showcases the effectiveness of our hierarchical framework in extremely large-scale reconstruction and rendering tasks.

\subsection{Implementation Details}
We implemented our Pyramid-GS framework in PyTorch. 
All experiments were conducted on a single NVIDIA RTX A6000 GPU with CUDA 11.6, and models were trained for 40,000 iterations using the Adam optimizer. 
For our Pyramid sampling, the number of levels $L$ is automatically set according to the dataset.
Regarding the Scene Perception Compensation module, the depth of field standard threshold $\sigma_{z,\text{thresh}}$ is set to 50.0, with a corresponding distance compensation coefficient $\alpha$ of 0.7. The coverage weight $\beta$ is 0.5. 
Finally, for our Laplacian Mixed Model, the rate-distortion trade-off hyperparameter $\lambda$ from Eq.~\ref{eq:loss_function} is uniformly set to 0.0005 for all reported results

\begin{figure*}[hp]
\centering
\includegraphics[width=0.95\textwidth]{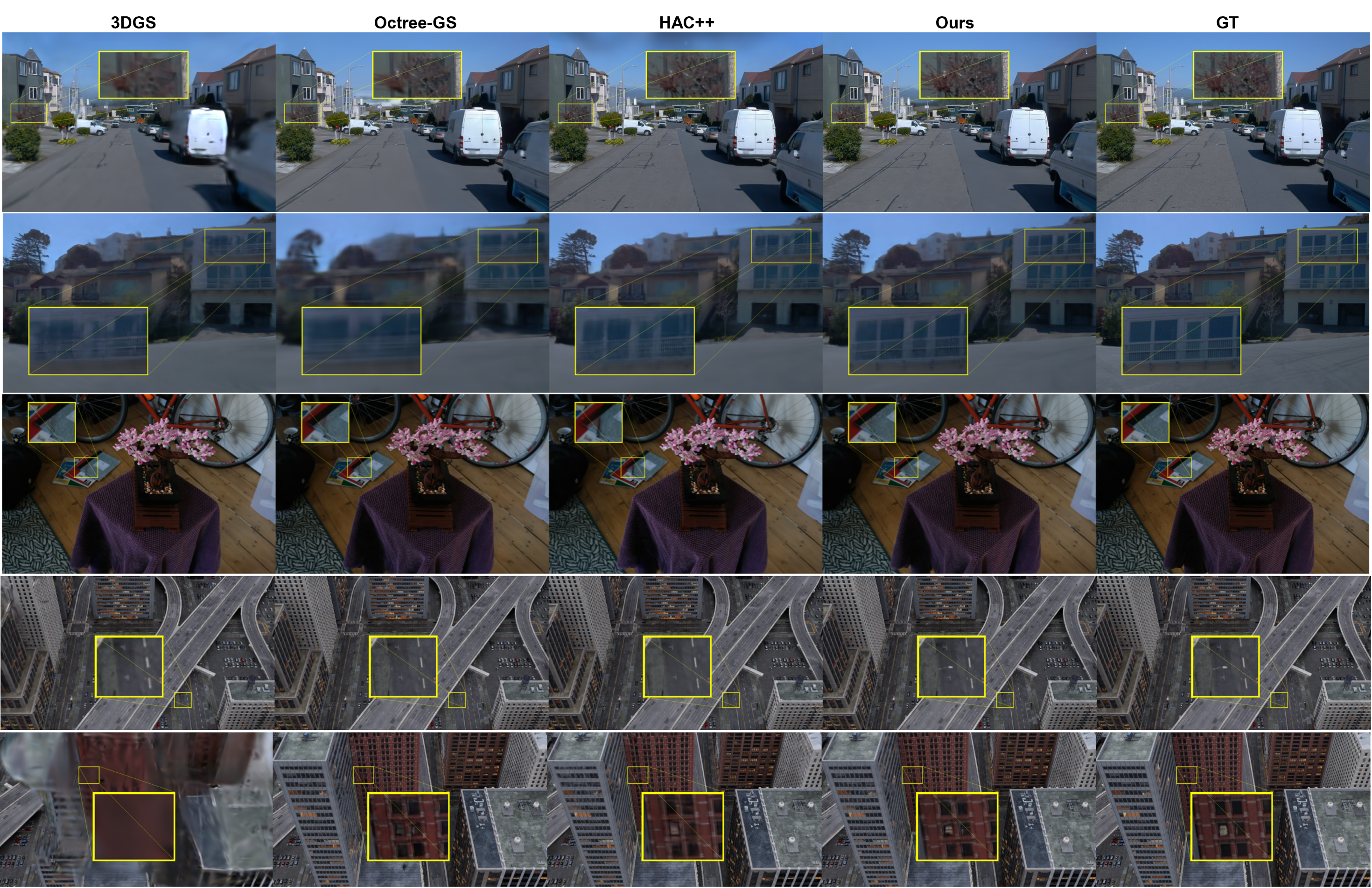}
\caption{\textbf{Comparison Results.} Visual differences are highlighted with yellow insets for better clarity. Our approach consistently outperforms other models on different scenes, demonstrating advantages in challenging scenarios. Best viewed in color.}
\label{r1}
\end{figure*}

\subsection{Comparison Experiments}
\label{subsec:comparison}

We conducted extensive comparison experiments against state-of-the-art 3D reconstruction and rendering methods, including 3DGS~\cite{kerbl20233d}, 2DGS~\cite{huang20242d}, Scaffold-GS~\cite{lu2024scaffold}, Hierarchical-GS~\cite{kerbl2024hierarchical}, Octree-GS~\cite{ren2024octree}, Context-GS~\cite{wang2024contextgs} and HAC++~\cite{chen2025hac++}. 
Our evaluation focuses on key metrics such as rendering speed, memory usage, primitive number and reconstruction quality (PSNR, SSIM~\cite{wang2004image}, LPIPS~\cite{zhang2018unreasonable}).

As demonstrated in Tab.~\ref{e1}, Tab.~\ref{e2} and Fig.~\ref{r1}, our Pyramid-GS consistently achieves competitive rendering quality and efficiency across a diverse range of datasets. 
These datasets span a wide range of scales and complexities, from object-centric small scenes to large-scale autonomous driving and massive urban environments, allowing for a comprehensive evaluation of the capabilities of our method.

Specifically, Tab.~\ref{e1} highlights the effectiveness of our Pyramid-GS on large-scale datasets. 
On Waymo dataset, our method achieved optimal rendering results with minimal memory consumption, underscoring the scalability and efficiency of our approach. 
For Matrix City, we achieved the best rendering quality, with memory consumption only slightly increasing compared to HAC++ and Context-GS. 
These results confirm the advantage of our hierarchical scene perception representation in significantly reducing the number of active Gaussians that need to be processed and stored, thereby improving memory efficiency on large scenes with the compression algorithm.

In contrast, Tab.~\ref{e2} presents results for small-scale scenes, such as Mip-NeRF360 and Tanks $\&$ Temples. 
On Mip-NeRF360, our method achieved the most best results, while on Tanks $\&$ Temples, we reached a near-optimal state, notably utilizing the fewest Gaussians to achieve this rendering quality. 
This is because the Tanks $\&$ Temples are dense and involve more complex textures, while our model pays more attention to large-scale and efficient management than the rendering quality, so it can achieve suboptimal rendering quality with the least number of Gaussians.
This reinforces our core contribution: a novel scene perception and compression framework for 3D Gaussian Splatting.

The experimental results reveal that our Pyramid-GS excels at maintaining high rendering quality across diverse scales due to our scene perception compensation algorithm. 
Furthermore, our pyramid representation and compression module fosters a flexible and memory-efficient scene representation. 
It allows for adaptive rendering that scales gracefully with scene complexity, preserving essential details even at lower levels of the hierarchy. 
We present qualitative results in Fig.~\ref{r1}, which show that our model maintains photorealistic rendering quality and handles complex geometry effectively across different scales. 
The performance gains are more significant in large-scale environments, confirming the robustness of our perceive-sample-compress approach.

\begin{figure}[]
\centering
\includegraphics[width=0.44\textwidth]{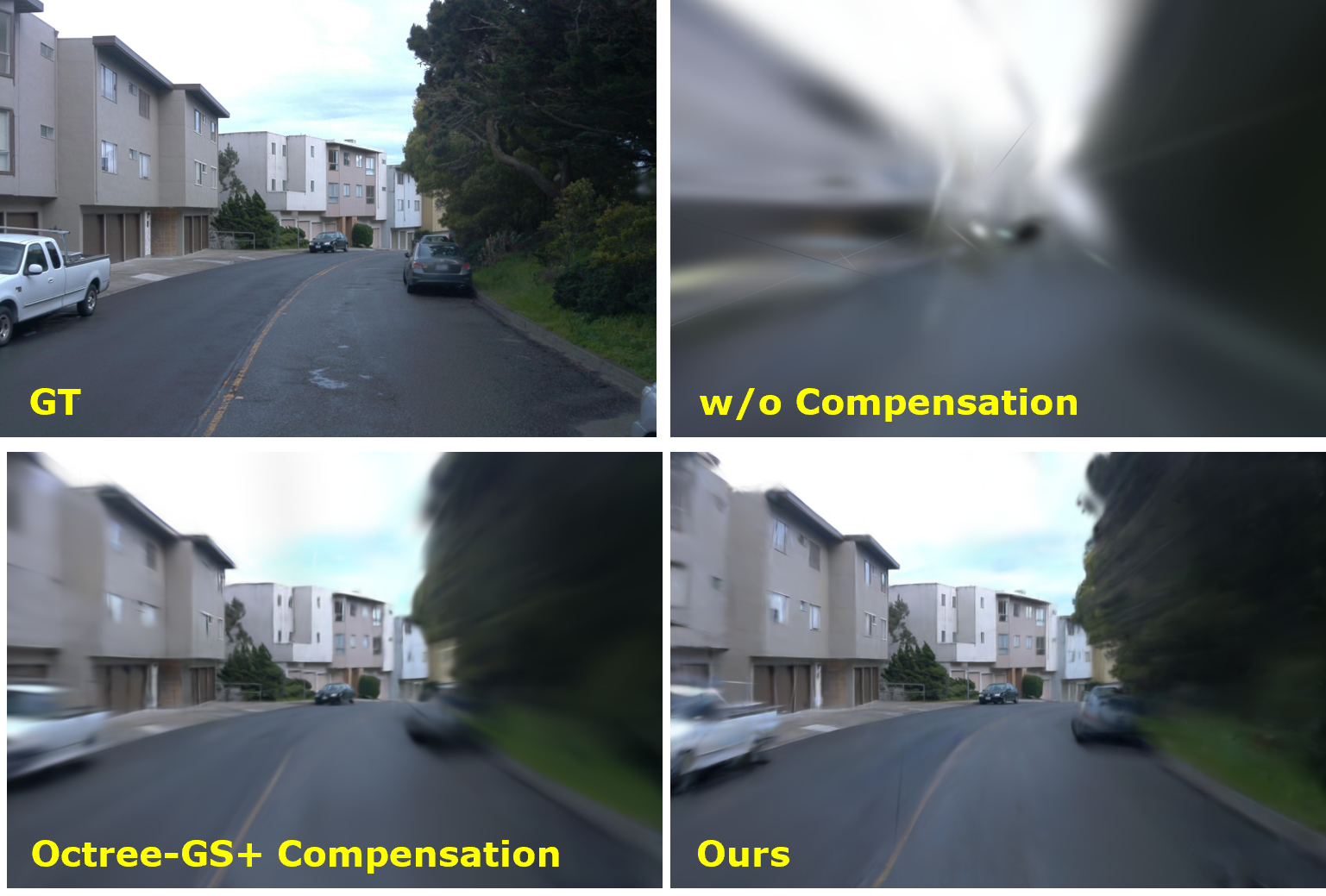}
\caption{\textbf{Effect of Scene Perception Compensation.} While relying solely on camera distance can work well in most situations, in some extreme scenes it may fail and result in poor rendering results. This is because that too many key points are ignored by the algorithm. Through our algorithm, the model can correct errors and preserve details. }
\label{compensation}
\end{figure}

\subsection{Ablation Studies}
\label{subsec:ablation}

To evaluate the contribution of each proposed component, we performed comprehensive ablation studies as shown in Tab.~\ref{ablation}, and Fig.~\ref{compensation}.

\paragraph{Impact of Scene Perception Compensation}
First, we evaluate the importance of the scene perception compensation algorithm. 
We compare the Pyramid-GS with a variant that uses a camera distance-based level partition and mask without considering the scene compensation.
The results in Tab~\ref{ablation} show that the disabling of scene perception compensation leads to a degradation in rendering quality. 
In certain complex scenes, relying solely on distance can cause important details to be oversimplified. 
The depth and coverage compensation ensures that Gaussians relevant to the viewer's perspective are prioritized, leading to better perceptual quality and more efficient rendering of visually significant features.
As shown in Fig.~\ref{compensation}, in some extreme situations, the camera distance-based function will lose control and fail to optimize the scene reconstruction.

\begin{table}[h]
  \centering
  \caption{Ablation study of different components on Waymo dataset, where SPC means the scene perception compensation algorithm, PS denotes pyramid sampling.}
  \label{tab:ablation_components}
  \resizebox{\columnwidth}{!}{%
    \begin{tabular}{l|ccccc}
      \toprule
        & SSIM↑ & PSNR↑ & LPIPS↓ & \#GS(k)↓ & Mem(MB)↓ \\
      \midrule
        w/o SPC & 0.833 & 27.12 & 0.312 & 138 & 14.5 \\
        w/o PS              & 0.845 & 27.95 & 0.301 & 154 & 17.2 \\
        w/o SPC $\&$ PS                 & 0.846 & 27.90 & 0.308 & 115  & 11.3 \\
        w/o Compression                  & 0.847    & 27.96    & 0.301    & 144  & 81.3 \\
        \midrule
        Ours                            & \cellcolor[HTML]{F09BA0}0.851 & \cellcolor[HTML]{F09BA0}28.18 & \cellcolor[HTML]{F09BA0}0.300 & \cellcolor[HTML]{F09BA0}105  & \cellcolor[HTML]{F09BA0}9.5\\
      \bottomrule
    \end{tabular}
  }
\label{ablation}
\end{table}

\paragraph{Impact of Pyramid Sampling}
We evaluate the efficacy of our Laplacian pyramid-based hierarchical representation against a baseline approach that uses only voxelization without the multi-level aggregation.
As shown in Table~\ref{ablation}, the full Pyramid-GS with Laplacian pyramid representation significantly reduces the number of active Gaussians. 
This hierarchical structure not only decreases the number of primitive by about 30\% compared to simple voxelization but also leads to a noticeable improvement in reconstruction quality. 
This highlights the benefit of our multi-level aggregation strategy for efficient scene management.

When we replace the pyramid representation with the Octree-GS, we can observe a degradation in rendering quality, this is because we can avoid over-pruning important points through a backward upsampling process.

\begin{table}[h]
  \centering
  \caption{Ablation study on density threshold $\lambda$ on Waymo dataset. Increasing $\lambda$ reduces memory consumption and number of Gaussian primitive, but slightly affects quality.}
  \label{tab:ablation_lambda}
  \resizebox{\columnwidth}{!}{%
    \begin{tabular}{c|cccccc}
      \toprule
      \begin{tabular}[c]{@{}c@{}} $\lambda$\end{tabular} 
      & SSIM↑ & PSNR↑ & LPIPS↓ & \#GS(k)↓ & Mem(MB)↓ & FPS↑ \\
      \midrule
      0.0005 & 0.851 & 28.18 & 0.300 & 105 & 9.5 & 138 \\
      0.0014 & 0.848 & 28.07 & 0.310 & 81 & 6.9 & 152 \\
      0.0018 & 0.846 & 27.95 & 0.315 & 76 & 6.0 & 159 \\
      0.0030 & 0.842 & 27.80 & 0.322 & 66 & 5.1 & 161 \\
      0.0040 & 0.839 & 27.65 & 0.329 & 56 & 4.1 & 163 \\
      \bottomrule
    \end{tabular}
  }
\label{lambda}
\end{table}

\paragraph{Impact of Compression Algorithm}
Finally, we evaluate the effectiveness of our Generalized Gaussian Mixed compression algorithm for storing the Gaussian representations. 
Tab.~\ref{ablation} demonstrates that our compression algorithm can achieve significant size reductions with minimal impact on rendering quality. 
When replacing it with the Gaussian Mixed model used in HAC++, we can observe a degradation in both rendering quality and memory consumption.
This highlights the efficiency of our compression scheme, which leverages the structure of the pyramid to effectively represent Gaussian primitives with fewer parameters, making our Pyramid-GS highly practical for storage and deployment.

\begin{table}[h]
  \centering
  \caption{Comparison of training, encoding, and decoding time across different methods.}
  \label{tab:time_comparison}
  \resizebox{\columnwidth}{!}{%
    \begin{tabular}{l|ccc}
      \toprule
       Method & Training Time & Enc. Time(s) & Dec. Time(s) \\
      \midrule
      Octree-GS~\cite{ren2024octree} & 49min  & - & - \\
      Context-GS~\cite{wang2024contextgs} & 1h58min  & 78.39 & 72.59 \\
      HAC++~\cite{chen2025hac++}     & 1h34min  & 30.81 & 48.81 \\
      Ours      & 1h38min  & 16.31 & 28.95 \\
      \bottomrule
    \end{tabular}
  }
\label{time}
\end{table}

\subsection{Discussion}
We explore the relationship between the compression rate, rendering quality, speed, and final storage size by adjusting the $\lambda$. 
Tab.~\ref{lambda} shows that increasing $\lambda$ results in higher compression rates, which reduces the final storage size and increase the FPS. 
However, this also leads to a decrease in rendering quality. 
These findings highlight the trade-offs involved: higher compression leads to more compact storage and faster rendering but at the expense of visual fidelity. 

Additionally, we observe that our method has comparable enc/dec times for compression and overall training times to HAC++, indicating similar computational efficiency in Tab.~\ref{time}. 
This demonstrates that our approach maintains competitive performance while achieving effective compression.
As for the training time, it is longer than non-compression methods.
This is because joint optimization takes more time.

\subsection{Conclusion}

In this paper, we proposed Pyramid-GS, a novel approach to 3D Gaussian Splatting that leverages a hierarchical scene perception and compression representation for efficient primitive management. 
Through extensive experiments across diverse datasets, we demonstrated the effectiveness and scalability of our method in both small-scale and large-scale environments. 
Our ablation studies further validated the contributions of key components, emphasizing the importance of scene perception compensation, pyramid sampling and compression algorithm. 
The results indicate that Pyramid-GS not only enhances rendering quality but also significantly improves computational efficiency, making it a valuable tool for various applications in 3D reconstruction.

\bibliography{aaai2026}

\end{document}